\documentclass[journal=aelccp,manuscript=article]{achemso}
\usepackage{color}
\usepackage{tikz}
\usepackage{fontawesome}

\usepackage[version=3]{mhchem} 
\usepackage{pdfpages}

\author{Vikram Pande}
\affiliation{Department of Mechanical Engineering, Carnegie Mellon University, Pittsburgh, Pennsylvania 15213, USA}
\author{Venkatasubramanian Viswanathan}
\affiliation{Department of Mechanical Engineering, Carnegie Mellon University, Pittsburgh, Pennsylvania 15213, USA}
\email{venkvis@cmu.edu}

\title{Computational Screening of Current Collectors for Enabling Anode-free Lithium Metal Batteries}

\begin{document}

\begin{tocentry}
\begin{center}
\includegraphics[width=7cm]{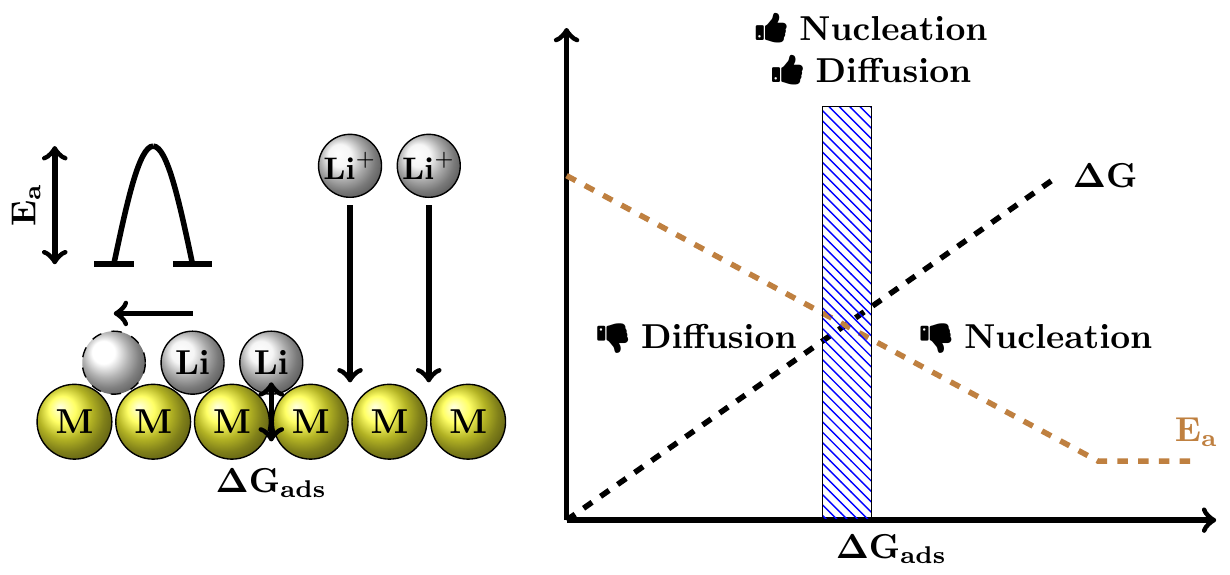}
\end{center}
\end{tocentry}

\begin{abstract}
Lithium metal cells are key towards achieving high specific energy and energy density for electrification of transportation and aviation.  Anode-free cells are the limiting case of lithium metal cells involving no excess lithium and the highest possible specific energy. In addition, anode-free cells are easier, cheaper and safer as they avoid handling and manufacturing of lithium metal foils. Issues related to dendrite growth and poor cycling are magnified in anode-free cells due to lack of excess lithium. Electrolyte and current collector surface play a crucial role in affecting the cycling performance of anode-free cells. In this work, we have computationally screened for candidate current collectors that can nucleate lithium effectively and allow uniform growth.  These are determined by the free energy of lithium adsorption and lithium surface diffusion barrier on candidate current collectors. Using density functional theory calculations, we show that Li-alloys possess ideal characteristics for Li nucleation and growth.  These can lead to vastly improved specific energy compared to current transition metal current collectors.
\end{abstract}

\newpage
Lithium metal cells are key towards achieving high specific energy and energy density energy storage to enable electrification of transport\cite{choi2016promise,unitedusabc,sripad2017performance,xu2014lithium} and aviation\cite{fredericks2018performance,viswanathan2019potential,hepperle2012electric}. Most of the current research on lithium metal cells use a lithium foil at the anode (of varying thickness)\cite{park2014highly,li2016artificial}. Given the cathodes are already pre-lithiated, this excess lithium results in a lower energy density than theoretical limit, but improves cycle life by increasing lithium inventory.\cite{jiao2018stable,zhang2017lithium,lin2017reviving,zhang2018ionic,fan2018non} Anode-free cells are the limiting case of lithium metal cells involving no excess lithium and thus, the highest possible energy density.\cite{louli2019exploring,qian2016anode,neudecker2000lithium} Anode free cells are made out of a fully lithiated cathode stacked with the separator and current collector as shown in SI, Fig. S1. During the first charge, the lithium stored in the cathode is deposited on the current collector as metallic lithium and then intercalated in the cathode at subsequent discharge.\cite{kerman2017practical} Anode free cells are also easy and safe to construct as they avoid handling and manufacturing of lithium metal foils.\cite{qian2016anode} In addition, high-quality thin lithium foils are expensive and one of the major economic risks associated with practical lithium metal batteries.\cite{albertus2018status}  An anode-free design circumvents this issue and thus, can enable both easily manufacturable and cost-competitive lithium metal batteries.

Lithium metal cells using liquid electrolytes are limited by low coulombic efficiency and dendrite growth.\cite{lin2017reviving,zhang2017lithium,lin2016layered,wood2017lithium} These problems are significantly magnified in anode free cells due to the lack of excess lithium.\cite{louli2019exploring,genovese2018combinatorial,qian2016anode} The large volume expansion of the plated lithium during cycling in anode free cells leads to a large stress on the SEI resulting in cracking and thus exposing more lithium to the electrolyte for furthur parasitic reactions. Another important difference in anode free cells is that the lithium nucleation occurs on the current collector surface, significantly different from nucleation on lithium itself. This can lead to nucleation overpotential losses and also affect lithium deposition morphology resulting in dendrite formation.\cite{pei2017nanoscale} Modifications to the copper current collector surface have already shown improvement in coulombic efficiency and compact lithium deposition.\cite{lu2016free,zhao2018compact,yan2018three} A variety of coatings such as transition metals and carbon/graphene on copper have also been used to modify the lithium nucleation and in turn the morphology.\cite{assegie2018polyethylene,liu2018dendrite,zhang2017lithiophilic,chi2017prestoring,yan2016selective} Pei et. al. have also shown the dependence of lithium nuclei size, shape and areal density on the applied current density.\cite{pei2017nanoscale} In general, larger current density results in larger number of nuclei with small sizes. This will result in increased surface area and in turn increased SEI formation and poor coulombic efficiency. Thus, in addition to identifying electrolytes that can lead to high-performing SEIs,\cite{zhu2019design,chiang2017lithium,qian2015high,zheng2017electrolyte,zhang2017fluoroethylene} designing the current collector surface becomes equally important for anode-free batteries.   

In this work, we focus on developing a detailed understanding of lithium nucleation and growth on a variety of candidate current collectors using density functional theory calculations, taking an electrocatalytic perspective on the problem.\cite{Exner2019RecentPlots}  Using a thermodynamic analysis based on the density functional theory calculations, we determine the thermodynamic nucleation potential and Li surface diffusion activation energies. We find that Li alloys are much better candidates as current collectors compared to the transition metals as they have very good Li nucleation and Li surface diffusion. We found a correlation between Li adsorption energy and Li diffusion activation energy. This relationship clearly shows that the best performing current collector surfaces should possess Li adsorption energy close to zero. 

\begin{figure}
    \centering
    \includegraphics[width=0.75\textwidth]{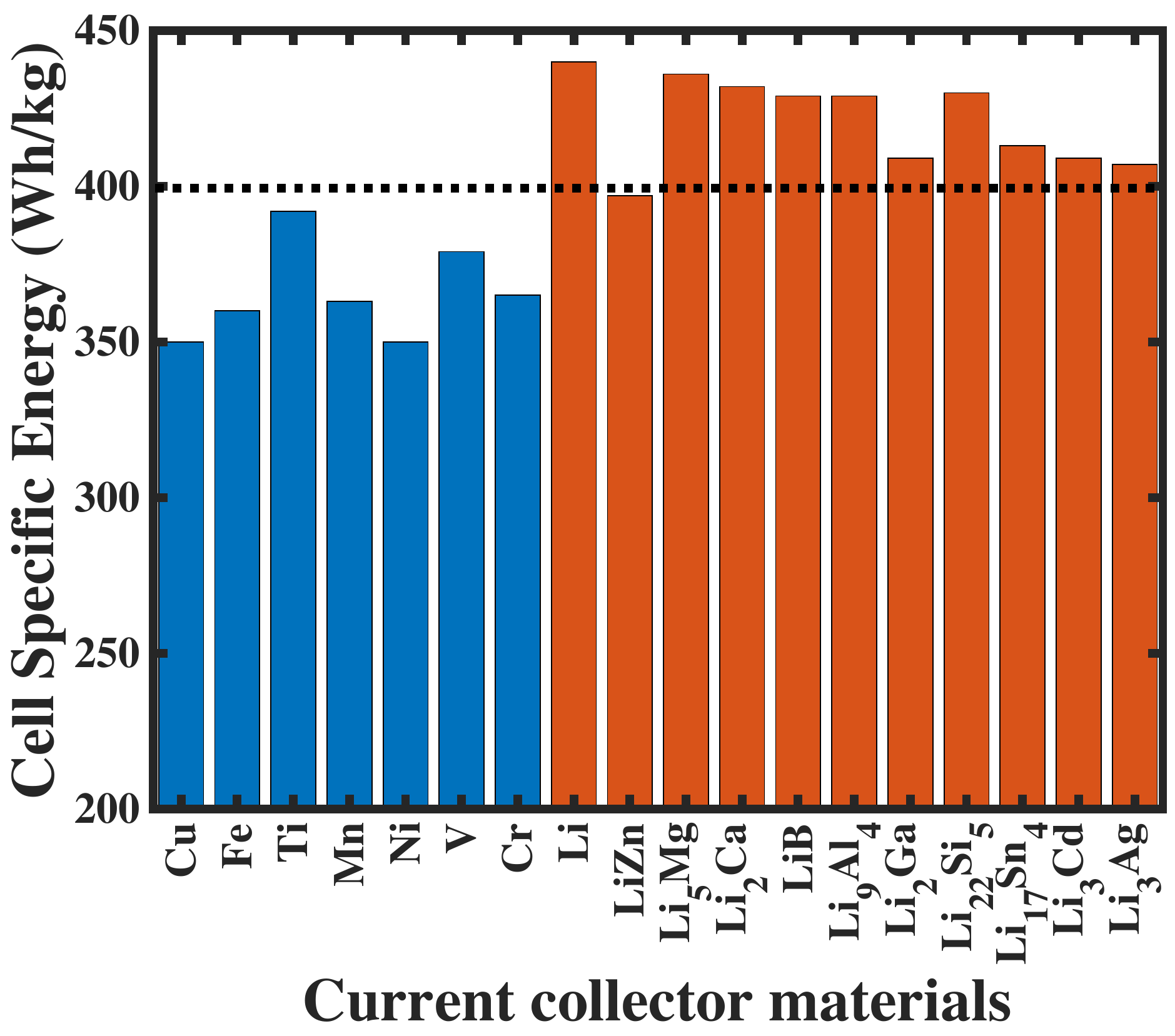}
    \caption{The specific energy of anode free cells using $10~\mu m$ current collectors made of different transition metals and alloys. Note that all these cells would have the same energy density. The cell design is taken from the work done by Zhu et al.\cite{zhu2019design} The cell comprises of a 4.25 mAh/cm$^2$ LCO cathode with 60$\mu$m thickness, 25$\mu$m thick separator, 15$\mu$m Al current collector at the cathode and 10$\mu$m anode current collector. Note there is no Li in anode free batteries.}
    \label{ccse}
\end{figure}

There are two possible approaches for an anode-free design: (i) replace copper as current collector completely or (ii) apply a coating of material on top of copper. As shown in Fig. \ref{ccse}, the first approach of replacing copper will lead to additional benefit of increasing energy density even further. This is largely attributed to the high density of Cu (8.96 g/cc) compared to the proposed candidates and lithium (0.5 g/cc). Specifically an anode free configuration with Li-alloys will give a specific energy $>$ 400 Wh/kg compared to 350 Wh/kg with Cu. Use of coatings will not affect the specific energy as Cu is still used. Thus, we focus from here on will be to identify better current collector candidates compared to Cu. 

A material must possess the following necessary properties, in addition to others, for use as a current collector in anode free batteries:
\begin{itemize}
    \item High electronic conductivity
    \item Stable against corrosion
    \item Li nucleation potential leading to 2D growth
    \item Fast surface diffusion of Li on the surface
\end{itemize}
The high electronic conductivity constraint restricts our search to metals and Li-alloys. Now additionally considering cost and abundance, the list of materials narrows down to Na, K, Cu, Fe, Ti, Ni, Cr, V, Mo, W, Zr, Mn as the transition metal elements and Li-Zn, Li-Al, Li-Ga, Li-B, Li-Si, Li-Sn, Li-Pb, Li-Cd, Li-Mg, Li-Ca, Li-Sr and Li-Ag. During operation of anode free batteries, the anode potential will likely be $\sim$ 0 V on the Li/Li$^+$ scale.\cite{monroe2003dendrite} The redox potentials of Ca, Sr and K is close to the anode potential implying that they may dissolve under these conditions. Na and Mg are highly reactive chemically and thus not considered. For the alloy materials, only the fully lithiated phases are considered as any other phase would consume lithium inventory during cycling. Thus the final list of materials considered is Cu, Fe, Ti, Ni, Cr, V, Mo, W, Zr, Mn, LiZn, Li$_9$Al$_4$, Li$_2$Ga, LiB, Li$_{22}$Si$_5$, Li$_{17}$Sn$_4$, Li$_{22}$Pb$_5$, Li$_3$Cd and Li$_3$Ag. Density functional theory (DFT) calculations were performed on the low miller index surfaces of all these materials to evaluate the Li nucleation overpotential and Li surface diffusion energy barrier. 

Self-Consistent DFT calculations were performed using the real space projector-augmented wave method implemented in the GPAW code.\cite{mortensen2005real,enkovaara2010electronic}. The Bayesian Error Estimation Functional with van der Waals (BEEF-vdW) exchange correlation functional was used for all adsorption free energy calculations owing to its accuracy for describing adsorption energies and energy barriers.\cite{wellendorff2012density,mallikarjun2017sbh10} For all calculations, the two bottom layers of the unit cell were constrained and the top two layers along with the adsorbates were allowed to relax with a force criterion of $<$ 0.05 eV/\AA. A Fermi smearing of 0.1 eV was used. The Brillouin zone was sampled using the Monkhorst Pack scheme and a k-point grid was chosen such that the $k_x L_x, k_y L_y, k_z,L_z > 40 \AA^{-1}$ where $k_x,k_y,k_z$ are the number of k-points and $L_x,L_y,L_z$ are the lengths of the unit cell in the x,y,z directions. To evaluate the nucleation overpotentials, we simulated a low coverage ($\theta$ $<$ 0.2) and the fully (1 ML) covered ($\theta$ = 1) surfaces. 

\begin{figure}
    \includegraphics[scale=0.21]{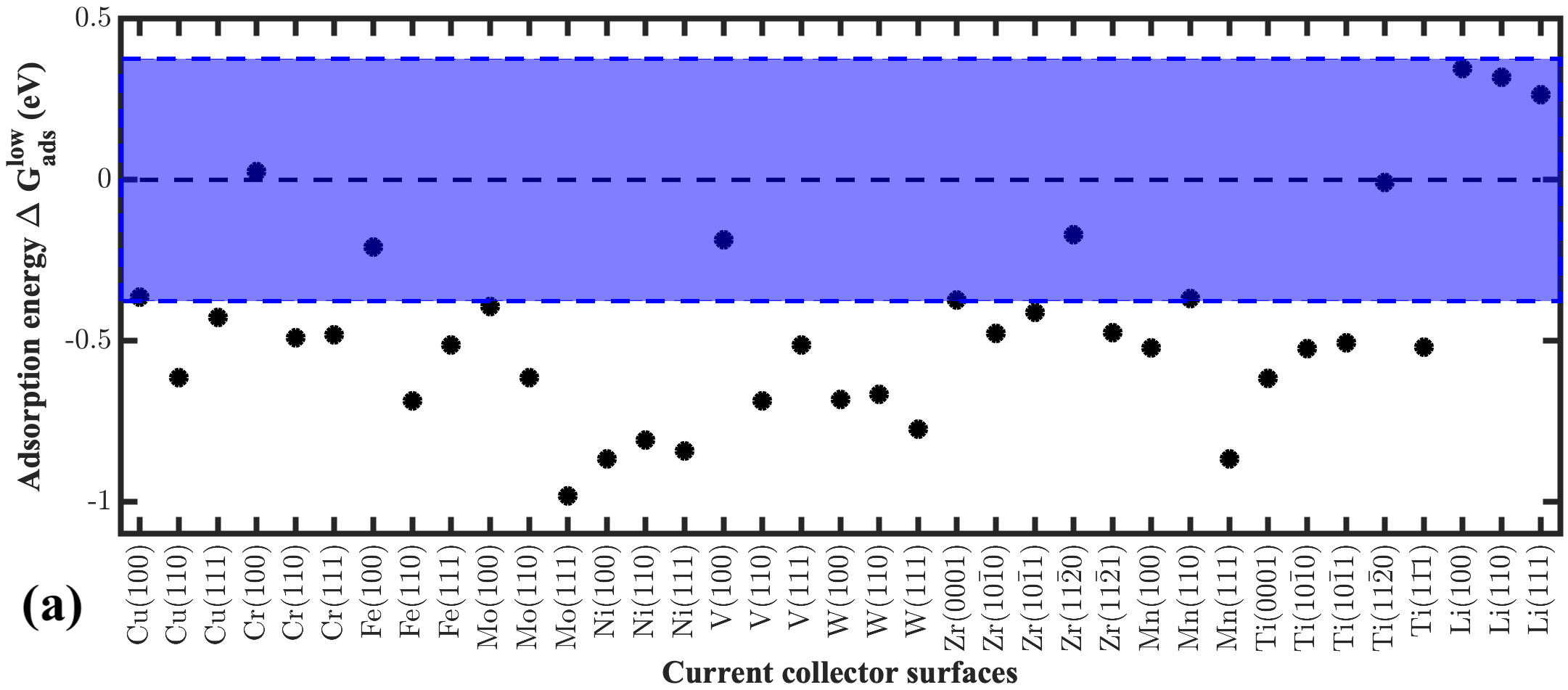}
    \includegraphics[scale=0.21]{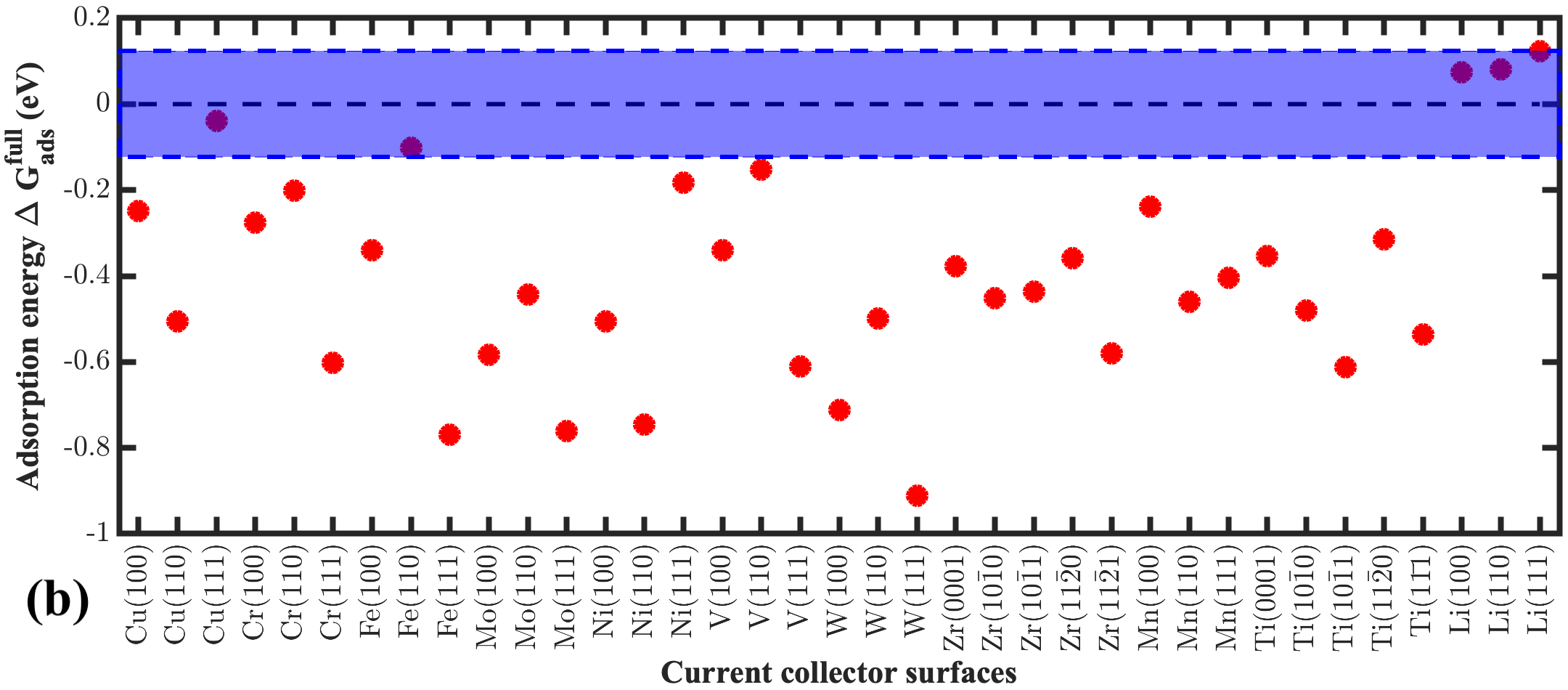}
    \includegraphics[scale=0.21]{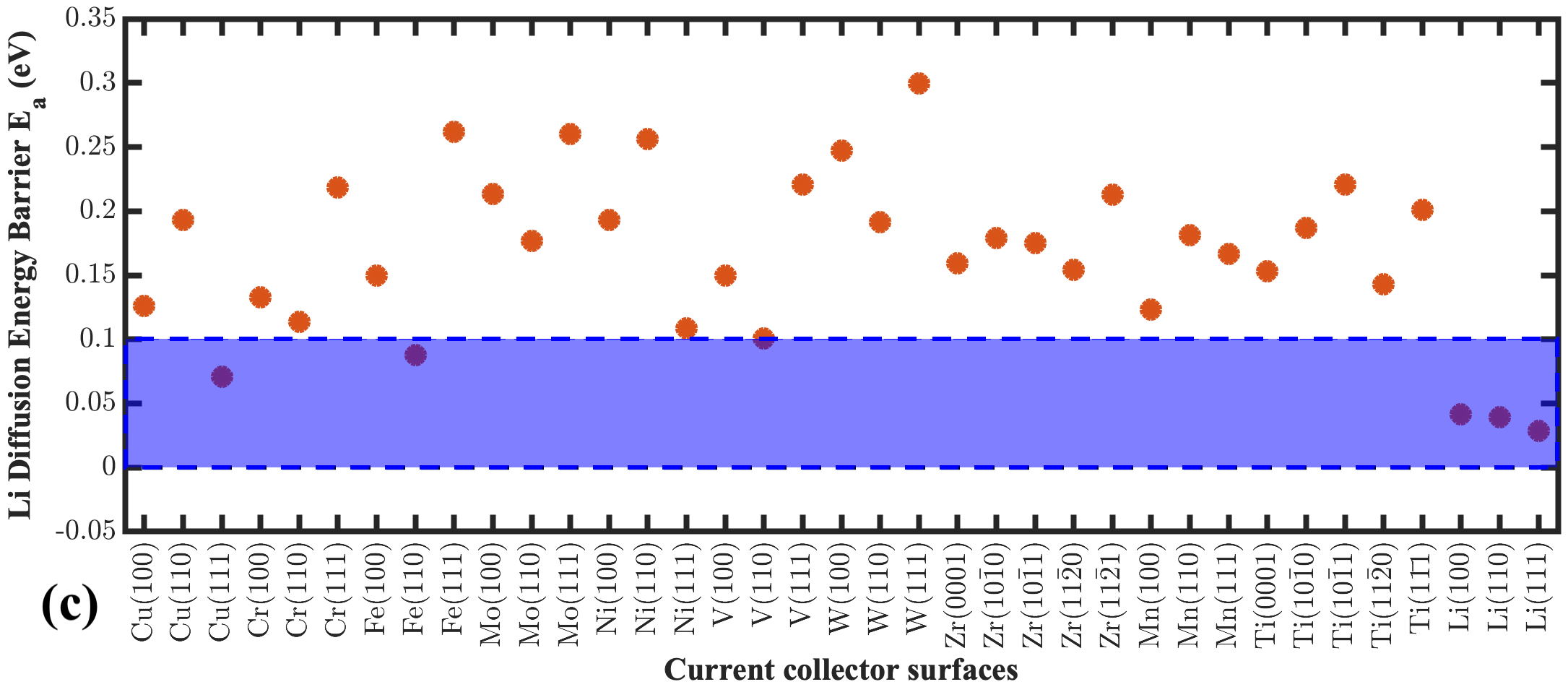}
    \caption{Li adsorption energies at (a) Low coverage, (b) 1 ML coverage on transition metal not forming an alloy with Li. (c) Li surface diffusion activation energies on transition metals. The blue regions denote the desired range for the properties.}
    \label{mlowads}
\end{figure}

At low Li coverage, we find that the Li nucleation overpotential on Li itself is about 0.3 V, while at 1 ML coverage it drops down to 0.1 V. We observe that most transition metals bind Li too strongly with an overpotential $>$ 0.3 V at low coverage as shown in Fig. \ref{mlowads} (a). Interestingly we find that Cr(100), Fe(100), V(100), Zr(11$\bar{2}$0), Ti(11$\bar{2}$0) and Mn(110) adsorb Li at low coverage with lower nucleation overpotential than Li itself. For Cr, Fe and V which are bcc crystals, we find that the Li is adsorbed in the hollow site and that the (100) facet has the weakest binding due to a higher coordination number of the surface atoms. Similarly for hcp metals Zr and Ti the weakest binding is for the (11$\bar{2}$0) surface and for Mn it is the (110) surface. 

At 1 ML Li coverage, almost all transition metal surfaces significantly over-bind Li with the exception of Cu(111), Fe(110), V(110) and Ni(111) as shown in Fig. \ref{mlowads} (b). For the (fcc) metals such as Cu and Ni, the Li atoms adsorb weaker on (111) surface compared to the (100) and (110) surfaces. This is because the Li coordination is 3 for (111) and 4 in the case of (100) and (110). For the bcc metals as Fe, Cr, Mo, etc, Li atoms adsorb the weakest on (110) facet again due to lower coordination. Lastly for the hcp crystals such as Zr and Ti, the (11$\bar{2}$0) facet has the weakest Li adsorption. Cu(111) has an exceptionally low nucleation at 1 ML coverage probably because of low coordination and similar lattice constants of Cu and Li.

%

From the surface energies given in SI Table S1, we see that all low index surfaces of Li have very similar surfaces energies. Thus, the nucleation overpotential will be governed by the best of the three surfaces and hence would be around 0.26 V for low coverage of Li and 0.07 V for 1 ML covered Li surface. For Cu, the (111) has the lowest surface energy and has very low 1 ML coverage overpotential but significantly high low coverage nucleation overpotential. Thus, twe find that increasing the fraction of (111) facet on the surface can reduce the overpotential. (111) is the most stable surface for Fe but (110) has a very good Li nucleation at 1 ML coverage. So Fe could potentially be used by increasing (110) fraction but this would be challenging due to thermodynamic stability. Similarly for V, the (111) surface is the most stable, but (100) and (110) surfaces have better Li adsorption characteristics. For Ni, (111) is the most stable surface and has moderate binding at 1ML coverage but over binds Li at low coverage. Ni can be used instead of Cu but would not provide any significant improvement. Among the transition metals considered, we do not find any candidates that would provide a good Li nucleation at both low and high Li coverage. So we believe that Li nucleation at best would be similar to Cu, which is the currently used current collector and provides inadequate performance. 

Out of all the Li-alloy surfaces, we observe that the Li-rich terminations are thermodynamically stable due to the fact that Li has the least surface energy compared to other elements.\cite{vitos1998surface} This means that on Li-alloy surfaces, we are effectively nucleating Li on a strained Li surface. It is well known that the adsorption characteristics can be tuned depending on the strain of the surface.\cite{mavrikakis1998effect,khorshidi2018strain} So the Li nucleation overpotentials for these Li-alloy surfaces are expected to be closer to Li than in the case of other transition metals considered above. The surface energies for the low miller index facets for these alloys are given in SI Table S2. 

For LiZn we find that (100) and (110) surfaces have the lowest surface energy. For Li$_3$Cd, (100), (110) and (111) have comparable surface energies. For Li$_3$Ag, the (001), (100), (110) and (111) have similar surface energy while the (101) has a higher surface energy and would exist at a lower fraction on the surface. For Li$_2$Ga, (001), (100), (101) and (111) surfaces will dominate the surface. For Li$_9$Al$_4$, (010), (100), (101), (110) and (111) surfaces will exist on the surface of the alloys. Lastly for LiB, the (10$\bar{1}$0), (10$\bar{1}$1) and (11$\bar{2}$0) have the low surface energies. So for further analysis, only these surfaces will be considered. As mentioned before, the surface energies of these stable surfaces are close to the surface energies of the Li surfaces (within 0.4 J/m$^2$), proving that the stable surfaces are indeed Li-like.

\begin{figure}
    \includegraphics[scale=0.21]{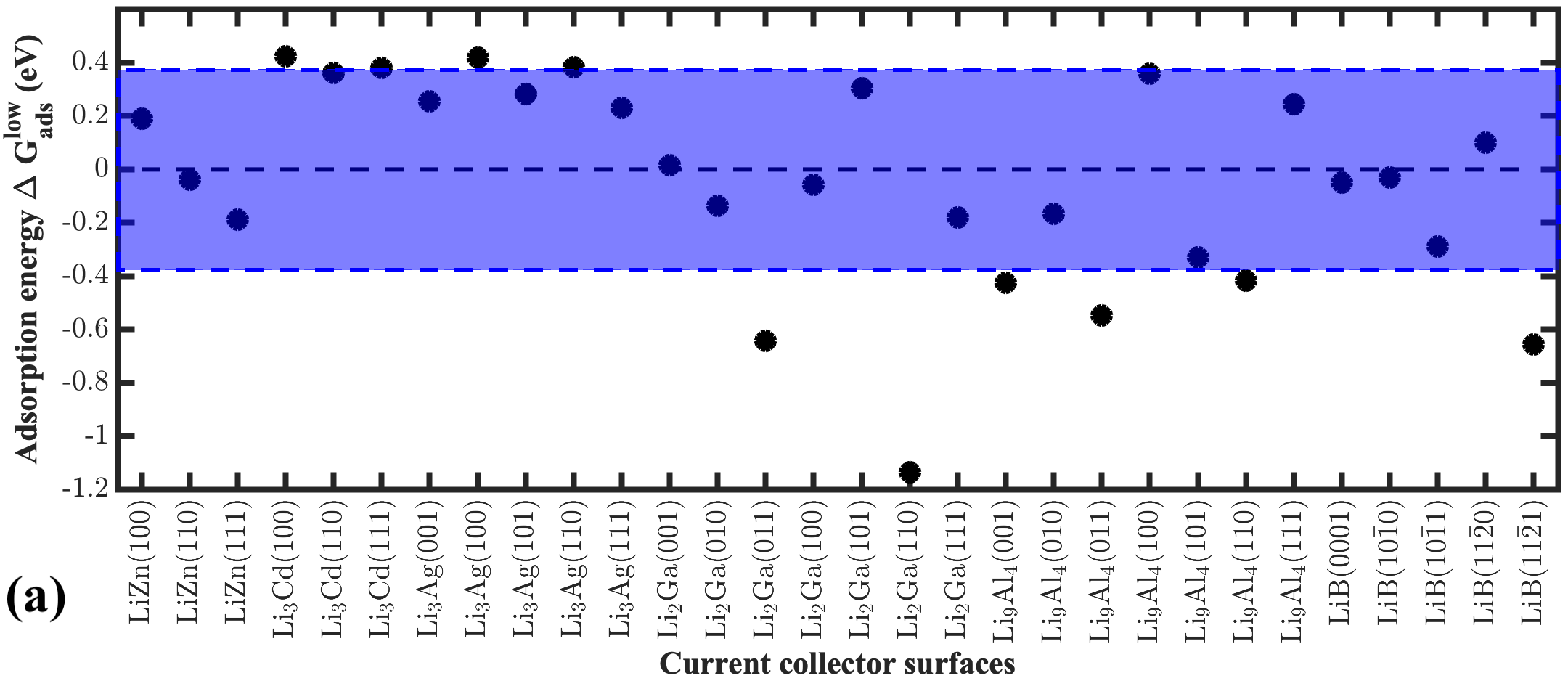}
    \includegraphics[scale=0.21]{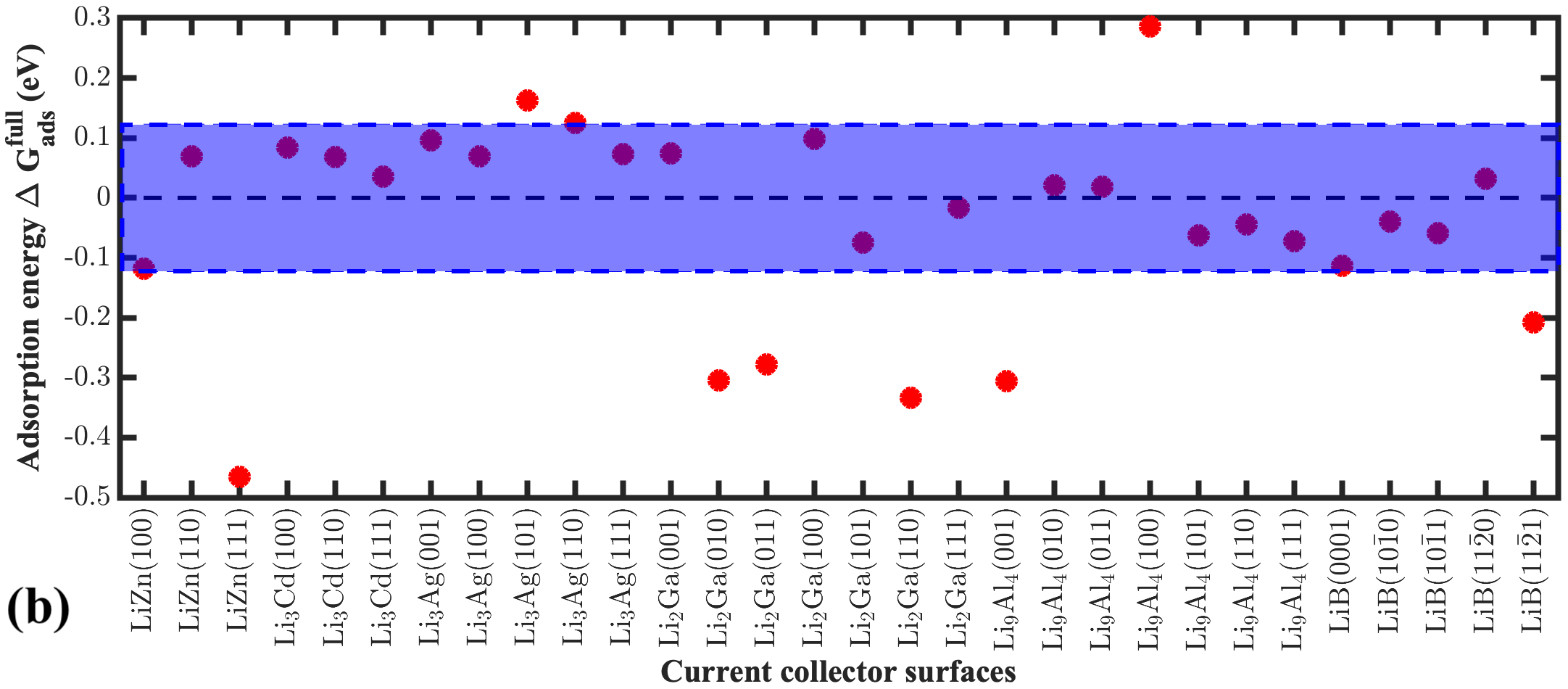}
    \includegraphics[scale=0.21]{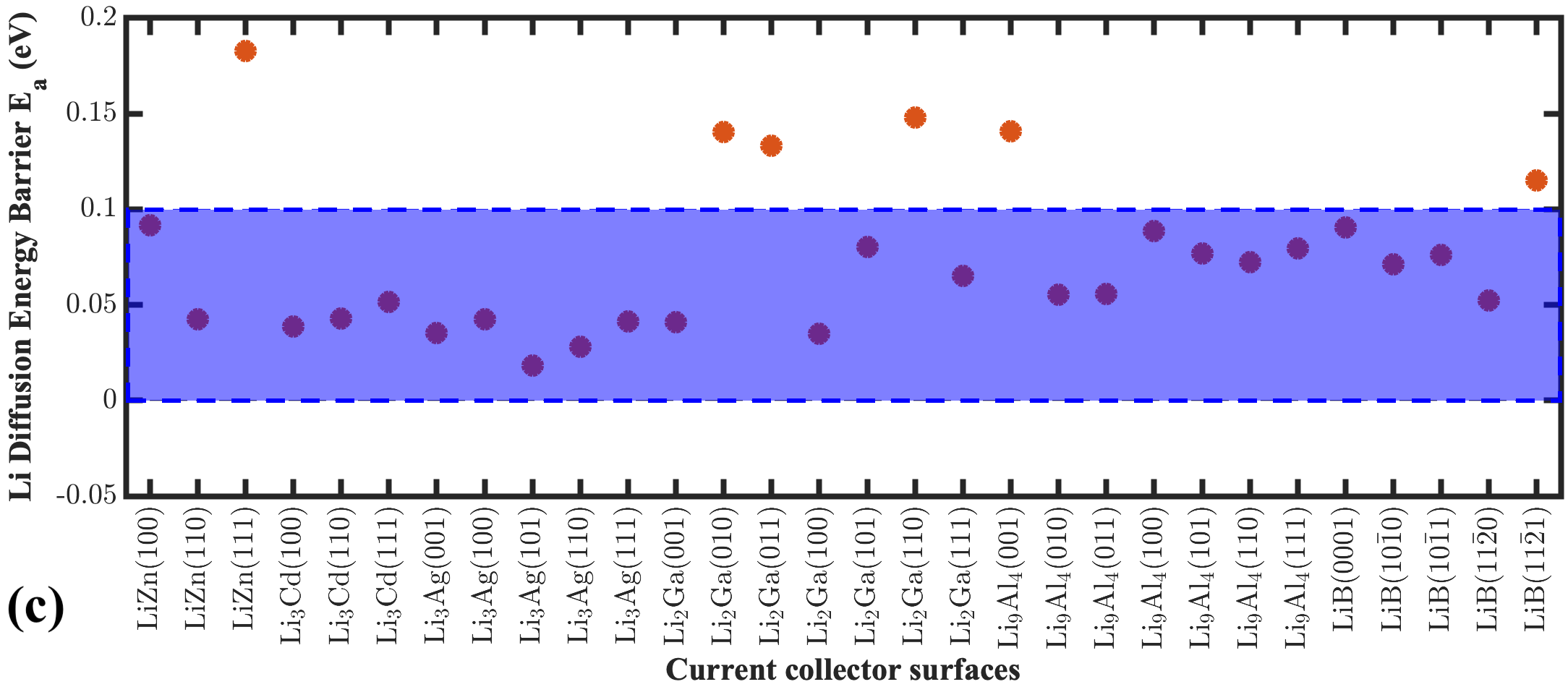}
    \caption{Li adsorption energies at (a) Low coverage, (b) 1 ML coverage on Li alloys. (c) Li surface diffusion activation energies on Li alloys. The blue regions denote the desired range for the properties.}
    \label{alowads}
\end{figure}

As shown in Fig. \ref{alowads} (a) for Li low coverage, Li$_3$Cd is slightly worse than Li. For Li$_3$Ag, (100) and (110) are similar to Li while (001) and (111) are significantly better. For Li$_2$Ga, (101) is similar to Li, (111) is slightly better but (001) and (100) have exceptionally low overpotentials. For Li$_9$Al$_4$, (010) and (100) are similar to Li, (011), (101) and (110) facets bind too strongly while (111) is better than Li. Lastly for LiB, (11$\bar{2}$0) facet has a non-existent overpotential, (10$\bar{1}$0) is good and (10$\bar{1}$1) is similar to Li.   

At 1 ML coverage for Li alloys, all the stable  surfaces for all Li-alloys have a nucleation overpotential lower than 0.1 V, which is the case for Li(111) as can be seen in Fig. \ref{alowads} (b). Interestingly, the 1 ML Li adsorption energy decreases with the surface energy of the Li alloy surface and increases as the strain on the Li monolayer with reference to the bulk increases as shown in SI Figs. S2 and S3. This clearly proves that the more Li-like the Li alloy surface, the better is the Li adsorption. Hence considering, nucleation overpotential losses, we clearly believe that Li alloys in many of the cases provide almost no nucleation overpotentials when compared to the standard transition metal current collectors.

Ensuring 2-dimensional growth at high rates, will depend on the surface diffusion of Li atoms on the current collector surface. It has been shown that fast surface diffusion can be used as a descriptor for uniform film growth.\cite{jackle2018self} During surface diffusion, the atoms jump from one site to the next site. The diffusion coefficient for such a process is given by\cite{oura2013surface}:
\begin{equation}
    D = D_0\exp\Bigg(-\frac{E_a}{k_BT}\Bigg)
\end{equation}
To a first approximation, we assume that the overall diffusion coefficient for Li diffusion on current collector surfaces is dependent on the activation energy. The Li surface diffusion activation energy was calculated using the nudged elastic band method\cite{jonsson1998classical} for 12 surfaces on the low coverage cases and the results are shown in Table. \ref{nebact}. Two adjacent adsorption sites were considered as the initial and final states for the surface diffusion calculation. The nudged elastic band method as implemented in atomic simulation environment\cite{larsen2017atomic} was employed to create five intermediate states for Li diffusion.

\begin{figure}
    \centering
    \includegraphics[scale=0.5]{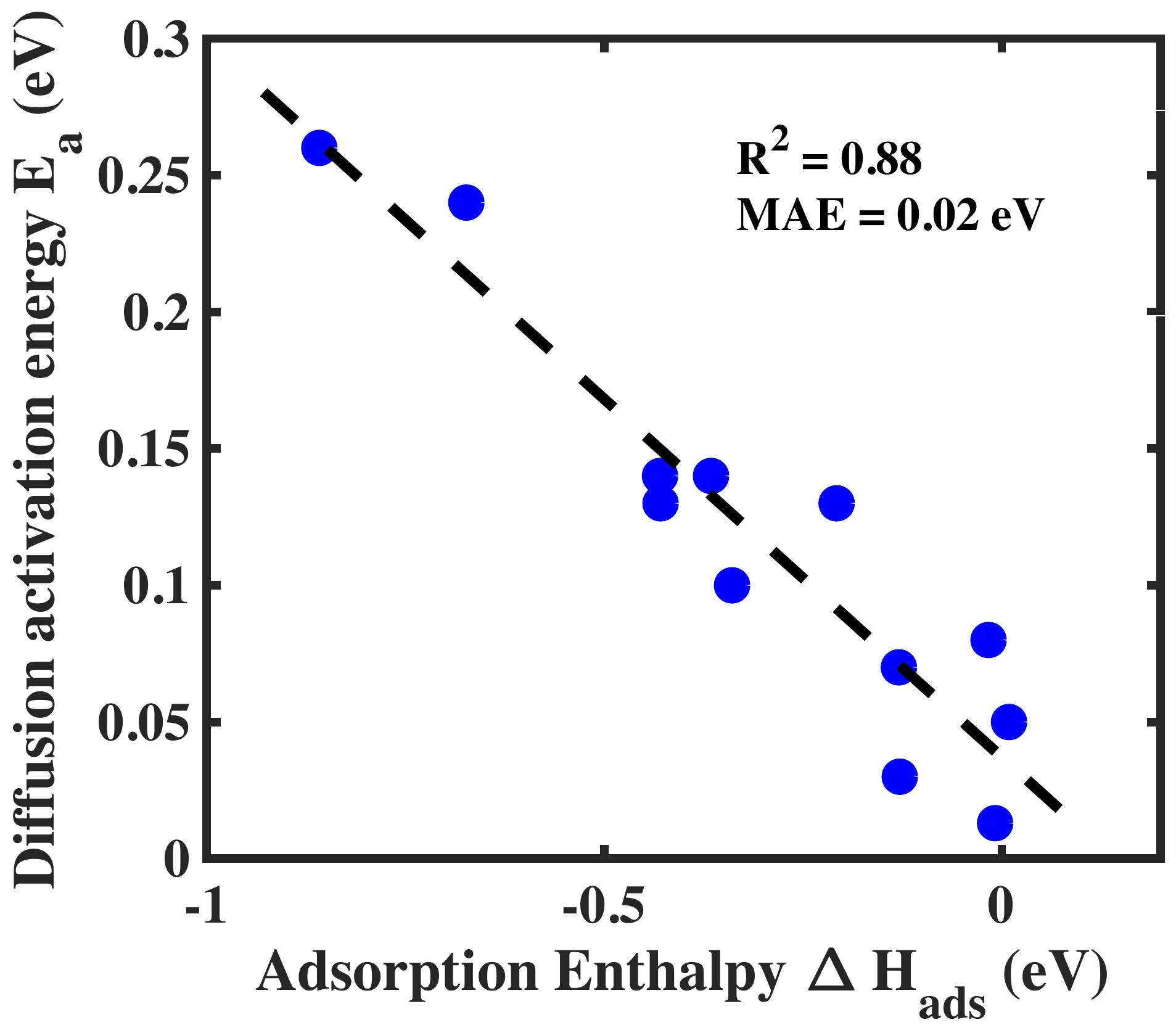}
    \caption[BEP relation for Li surface diffusion]{Br{\o}nsted–--Evans–--Polanyi (BEP) relation between adsoprtion enthalpy for nucleation for 1 ML Li coverage and the activation energy for  trained on 12 structures. The fit has an R$^2$ = 0.88 and MAE = 0.02 eV.}
    \label{beprel}
\end{figure}

To calculate the Li-diffusion activation energies for all the remaning surfaces, we derived a Br{\o}nsted---Evans---Polanyi (BEP)  relation\cite{evans1937introduction} between the activation energy and the adsorption enthalpy of 1 ML Li covered surfaces. BEP relations have been demonstrated for a variety of adsorbates on different transition metal surfaces and provide for an easy way to compute a large number of of activation energies.\cite{wang2011universal} NEB calculations are computationally expensive and thus, not feasible for determining diffusion activation energies for 70-80 structures. As expected, we find a strong correlation between the activation energy and the adsorption enthalpy of the 1 ML covered Li surfaces shown in Fig. \ref{beprel}. We find an excellent BEP relation with an MAE of 0.02 eV on the training set of activation energies. Now we use this derived relationship to determine the activation energy for all the remaining surfaces.

\begin{table}[!ht]
    \centering
    \begin{tabular}{|c|c|}
\hline
Surface	&	Activation Energy (E$_a$)	\\
\hline
Li(100)	&	0.08	\\
Li(110)	&	0.01	\\
Li$_2$Ga(100)	&	0.05	\\
Fe(100)	&	0.14	\\
Cr(100)	&	0.14	\\
V(100)	&	0.13	\\
Mo(100)	&	0.24	\\
Fe(111)	&	0.26	\\
Cu(100)	&	0.10	\\
LiZn(100)	&	0.13	\\
Cu(111)	&	0.03	\\
LiB(10$\bar{1}$0)	&	0.07	\\
\hline
    \end{tabular}
    \caption[NEB acitvation energies for used for BEP]{Activation energies calculate for a set of transition metal and Li alloy surfaces. }
    \label{nebact}
\end{table}

As given in Table \ref{nebact}, Li(110) has almost an ideal activation energy of 0.01 eV while Li(100) has a considerably larger value around 0.1 eV. Interestingly Cu(111) also has a very small barrier of about 0.03 eV. Considering Li(100) as benchmark, we consider materials that have an activation energy $<$ 0.15 eV to be good condidates as current collectors. Among the transition metal surfaces, Cu(111) has the lowest surface diffusion barrier of 0.03 eV. Other than that Cu(100), Cr(100), Cr(110), Fe(100), Ni(111), V(100), V(110), Zr(11$\bar{2}$0), Mn(100), Ti(0001) and Ti(11$\bar{2}$0) as seen in Fig. \ref{mlowads} (c), have sufficiently low activation energies. As mentioned before, out of these Cu(111), Ni(111) and Ti(0001) are thermodynamically stable and are probable candidates. However other may be used if grown epitaxially over other surfaces.

For all Li-alloy surfaces, except for LiZn(111) which is not thermodynamically stable, the activation energy is lower than defined criteria of 0.15 eV as shown in Fig. \ref{alowads} (c). So, all considered Li-alloys are good for Li surface diffusion as well. Out of all the alloy candidates, Li$_3$Ag(101) surface has the lowest barrier of 0.02 eV, while Li$_3$Ag(110) has a barrier of 0.03 eV. Considering the surface energetics, we find that all Li-alloys have average activation energies $\sim$ 0.05 eV. So on average, most of these should be better than Cu .

\begin{figure}
    \centering
    \includegraphics[scale=0.5]{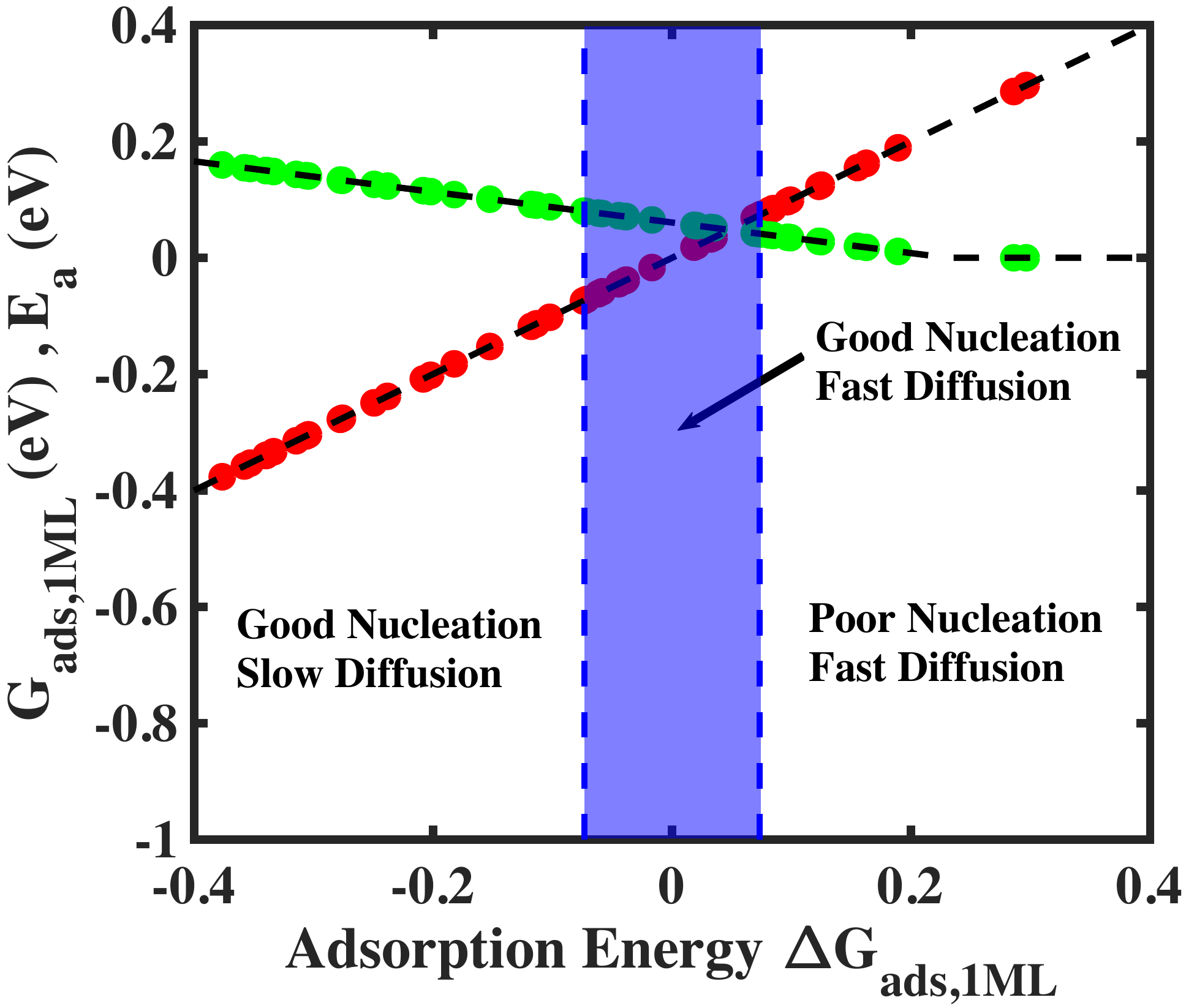}
    \caption{Volcano relationship for the performance of current collectors based on the single descriptor of 1 ML Li adsoprtion energy. The red dots are Li adsorption energies and green dots are the Li surface diffusion activation energies for all materials considered. The blue region depicts the region where the nucleation overpotential and activation energy is at least as good as on Li itself and thus optimal performance. }
    \label{ccvolcano}
\end{figure}

Fig. \ref{ccvolcano} shows that the 1 ML Li adsorption energy can be used as the descriptor for current collector performance. At low $\Delta G_{ads,1ML}$, Li binds strongly resulting in good nucleation but poor diffusion. At high $\Delta G_{ads,1ML}$, Li diffuses fast on the surface but does not nucleate. Thus there is a small optimal range where the nucleation overpotential is less than that of Li (100),  which will maximize performance.

Small diffusion activation energies in addition to slightly stronger binding on the Li-alloy surfaces in comparison to Li will also help in redistribution of the dendritic Li over time. Lastly it is well known in anode free batteries that at higher current rate, the Li nuclei size decreases and the nuclei number increases. This results in a tremendous increase in the surface area which results in significantly increased SEI formation reactions. Thus a decrease in coulombic efficiency is expected with increase in higher charging current. So we believe that Li-alloys with better nucleation and diffusion will improve performance at high charging rates for anode free cells.  

A few Li alloys have been experimentally tested in an anode-free configuration. Genovese et. al. have shown improvement in cycle life and lithium deposition morphology using Zn coating on a Ni current collector.\cite{genovese2018combinatorial} Yan et. al. have shown small Li nucleation overpotential on Au, Ag, Zn and Mg which form alloys with Li.\cite{yan2016selective} They used these metals in nanocapsules to get dendrite free Li deposition. Thus, Li alloys as current collectors would likely lead to a better nucleation of Li and improved cycle life of anode free batteries. However the effect of the current collector surface on cycle life will still be much smaller compared to the effect of the electrolyte additives.\cite{genovese2018combinatorial} For standard current collectors such as Cu and Ni we suggest surface modification that increases the fraction of (111) facet to improve Li deposition. It remains to be seen if the current collector surface is effective at suppressing dendrite growth under fast charging and low temperature conditions.

In summary, we have computationally screened for potential current collectors for anode-free lithium metal cells.  Using density functional theory calculations, we have calculated the nucleation overpotentials and surface diffusion activation energies for Li on various current collector material surfaces. Among the candidates considered, we find that using Li and Li-alloys as the current collector it is possible to develop cells with specific energy greater than 400 Wh/kg, which is challenging with standard transition metal current collectors such Cu, Ni and Ti. NEB calculations were done to derive a BEP relation, which was then used to determine the Li surface diffusion activation energies. Using the BEP relation, we show that to a first approximation, the 1 ML Li adsorption energy ($\Delta G_{ads,1ML}$) can be used as a descriptor for current collector performance, with optimal performance obtained when $\Delta G_{ads,1ML} \approx 0$. Li-alloys, Cu(111), Fe(110), V(110) and Ni(111) satisfy the above criterion. Thus, we propose the use of Li-alloys such as Li-Zn, Li-Al, Li-B, Li-Cd, Li-Ag, Li-Si, Li-Pb, Li-Sn, Li-Mg etc. as current collectors for anode free batteries to get high specific energies, low nucleation overpotentials, better rate capability and probably better control over dendrite in good electrolytes.



\begin{acknowledgement}
 Acknowledgment is  made  to  the  Scott Institute  for  Energy  Innovation at Carnegie Mellon for supporting V.P. during his graduate research. This work was supported by the Assistant Secretary for Energy Efficiency and Renewable Energy, Office of Vehicle Technologies of the US Department of Energy (DOE) through the Advanced Battery Materials Research (BMR) Program under contract no. DE-EE0007810.

\end{acknowledgement}

\noindent \textbf{Notes:} \\
V. P. and V. V. are inventors on US provisional patent application submitted by CMU that covers current collectors that enable lithium metal anodes.

\clearpage

\clearpage
\bibliography{refs.bib}

\includepdf[pages={1-}]{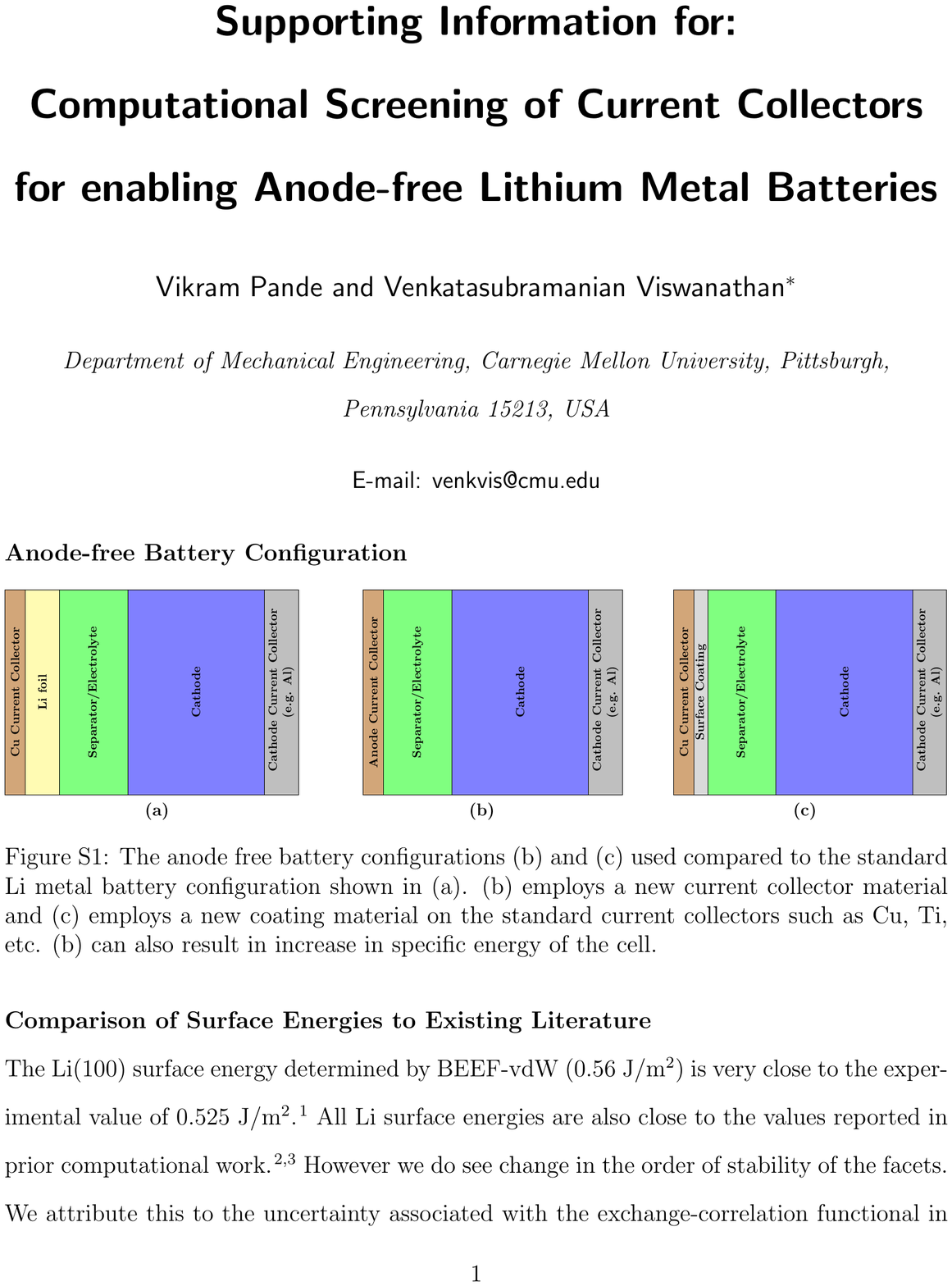}

\end{document}